*INVITED REVIEW PAPER*

# Review of Thermal Properties of Graphene and Few-Layer Graphene: Applications in Electronics

Zhong Yan[1], Denis L. Nika[1,2] and Alexander A. Balandin[1,3]

[1]Nano-Device Laboratory, Department of Electrical Engineering and Materials Science and Engineering Program, Bourns College of Engineering, University of California – Riverside, Riverside, California, 92521 USA

[2]E. Pokatilov Laboratory of Physics and Engineering of Nanomaterials, Department of Physics and Engineering, Moldova State University, Chisinau, MD-2009, Republic of Moldova

[3]Quantum Seed LLC, 1190 Columbia Avenue, Riverside, California 92507 USA

## Abstract

We review thermal properties of graphene and few-layer graphene, and discuss applications of these materials in thermal management of advanced electronics. The intrinsic thermal conductivity of graphene – among the highest of known materials – is dominated by phonons near the room temperature. The examples of thermal management applications include the few-layer graphene heat spreaders integrated near the heat generating areas of the high-power density transistors. It has been demonstrated that few-layer graphene heat spreaders can lower the hot-spot temperature during device operation resulting in improved performance and reliability of the devices.

*Keywords:* graphene, thermal conductivity, heat spreaders, thermal management

## I. Introduction

Thermal management represents a major challenge in the state-of-the-art electronics due to rapid increase of power densities [1, 2]. Efficient heat removal has become a critical issue for the performance and reliability of modern electronic, optoelectronic, photonic devices and systems. Development of the next generations of integrated circuits (ICs), high-power light-emitting diodes (LEDs), high-frequency high-power density communication devices makes the thermal management requirements extremely severe [1-8]. Theoretical and experimental studies have shown that micrometer or even nanometer scale hot spots generated in high-power density electronics due to non-uniform heat generation and heat dissipation may result in performance degradation and reliability issues [9, 10]. Efficient heat removal from hot spots area to the nearby surrounding area is the bottleneck of thermal management of high-power density devices such as GaN field-effect transistors (FETs). One possible approach for improving heat removal is introduction of micrometer or nanometer scale heat spreaders that specially designed for hot spots cooling. However, the thermal conductivity of semiconductor nanostructures is lower than that in corresponding bulk materials and diminishes with decreasing lateral dimensions [11-25]. This behavior of the thermal conductivity in nanostructures is explained by phonon boundary scattering and phonon confinement effects [18]. As a result, up to two-orders of magnitude reduction of the room-temperature (RT) thermal conductivity was reported for nanometer-thick films, multilayered planar heterostructures, homogeneous and segmented nanowires [19-25]. For this reason, the performance of nanometer scale heat spreaders implemented with conventional materials would be rather limited. Compared with metals or semiconductors, graphene has demonstrated extremely high intrinsic thermal conductivity, in the range from 2000 W/mK to 5000 W/mK at RT [26, 27].

This value is among the highest of known materials. Moreover, few-layer graphene (FLG) films with the thickness of a few nanometers also maintain rather high thermal conductivity unlike semiconductor or metals. Therefore, graphene and FLG are promising materials for micro or even nanometer scale heat spreader applications. Here we review the available thermal conductivity data for graphene and FLG and discuss their thermal applications as heat spreaders in high-power density electronics.

### II. Thermal properties of graphene and few-layer graphene

Graphene consists of a single layer of $sp^2$ bonded carbon atoms packed in honey comb lattice. In addition to its unique optical [28] and electronic properties [29, 30], graphene has demonstrated extremely high thermal conductivity [26, 27]. The thermal conductivity of graphene was measured, for the first time, at the University of California – Riverside using an optothermal method based on Raman spectroscopy [26, 27]. The development of the original *optothermal technique* was crucial for the thermal measurements of the atomically thin materials. In this technique, a Raman spectrometer acts as a thermometer measuring the local temperature rise in graphene in response to the Raman laser heating. Graphene has distinctive signatures in Raman spectra with clear G peak and 2D band [31, 32]. It has been demonstrated [33, 34] that the G peak position of graphene's Raman spectra exhibits strong temperature dependence. The shift in the position of the graphene G peak can be utilized to measure the local temperature rise.

Figure 1(a-b) shows the schematics of the optothermal measurement and a scanning electron microscopy of a representative suspended graphene flake used for the measurements [35]. A graphene layer is suspended across a trench fabricated on Si/SiO$_2$ wafer. The heating power $\Delta P$ is provided by the Raman excitation laser focused on a suspended graphene layer. Even a small amount of power dissipated in graphene can be sufficient for inducing a measurable shift in the G peak position due to the extremely small thickness of the material. The temperature rise $\Delta T$ in response to the dissipation power $\Delta P$ is determined through micro-Raman spectrometer. During the measurement the laser power is increased gradually and the local temperature rise in the suspended graphene layer is measured through $\Delta T = \Delta \omega_G / \gamma_G$ , where $\gamma_G$ is the temperature coefficient of graphene G peak. The amount of power dissipated in graphene layer can be determined via the integrated Raman intensity of G peak or by a power detector placed under the graphene layer. The correlation between $\Delta T$ and $\Delta P$ in graphene samples with given geometry can give the thermal conductivity values by solving heat diffusion equation.

Using the optothermal method, Balandin and co-workers found that the intrinsic thermal conductivity κ of suspended graphene flakes [26, 27] is in a range of 2000 W/mK – 5000 W/mK at RT, exceeding κ ~ 2000 W/mK of high quality bulk graphite. It has been established that the thermal conductivity varies with the size of graphene flake and that the heat is mainly conducted by acoustic phonons. The phonon mean free path (MFP) was estimated to be ~775 nm at RT [27]. The high values of thermal conductivity in graphene were confirmed by a number of independent experimental studies [36-38]. Cai et al. [36] performed measurements of the thermal conductivity of suspended high-quality chemical vapor deposited (CVD) graphene and found RT κ ~1500 - 5000 W/mK. A recent study used direct imaging of nanoscale thermal transport in single and few-layer graphene with ~50-nm lateral resolution using high vacuum scanning thermal microscopy [37]. It was concluded that the heat transport in suspended graphene is substantially increased as compared to the adjacent areas of supported graphene. The nano-thermal images indicated that the phonon MFP in the supported graphene can decrease down to 100 nm [37]. Another optothermal Raman study obtained a $\kappa$ value of ~1800 W/mK at 325 K and ~710 W/mK at 500 K [38] by assuming that the optical absorption coefficient is equal to 2.3% [39]. However, the absorption coefficient of graphene is a function of photon energy [40-42]. The value of 2.3% is observed only in the near-infrared at ~1 eV, and it steadily increases with increasing energy. Therefore value of 2.3 % assumed in Ref. [38] underestimated the amount of light absorbed by graphene, resulting in underestimation of the thermal conductivity. Optothermal method was also applied to measure thermal conductivity of few layer graphene (FLG) flakes of arbitrary shape via the iterative process of heat dissipation simulations [43]. Figure 2 shows the measured thermal conductivity of FLG as a function of number of graphene layers, $n$. It has been found that $K$ of suspended FLG will first decrease with increasing number of layers, then it will recover if $n$ keeps increasing and approaches bulk graphite limit ~2000 W/mK. This dependence of thermal conductivities on number of graphene layers can be explained by the intrinsic properties described by phonon-phonon scattering [43]. The increase in the number of graphene layers leads to the changes in the phonon dispersion, and results in more phonon states available for Umklapp scattering. The stronger Unklapp scattering decreases the intrinsic thermal conductivity.

Thermal conductivity of suspended graphene is closer to the intrinsic value which is determined by the phonon-phonon scattering. Thermal conductivity of the supported graphene is lower than that in suspended graphene due to the thermal coupling to the substrate and enhanced phonon scattering on the substrate defects and impurities. Seol et al. [44] have found $\kappa$~600 W/mK for graphene-on-SiO$_2$/Si near RT. Encasing graphene within two layers of SiO$_2$ leads to further reduction of the thermal conductivity $\kappa$ to value ~ 160 W/mK due to the phonon boundary and disorder scattering [45]. The thermal conductivity of the encased few-layer graphene increases with the increasing number of atomic planes and approaches the bulk graphite value. Thermal conductivities of graphene nanoribbons with less than five atomic layers and width between 16 nm and 52 nm were measured in the range $1000 - 1400$ W/mK using an electrical self-heating method [46]. The electrical breakdown current density was measured on the order of $10^8$ A/cm$^2$, close to that of carbon nanotubes (CNTs). High electrical breakdown current density [46, 47], along with the high thermal conductivity, suggest potential applications of graphene nanoribbons as interconnects in next generation ICs.

Theoretical investigations of the lattice thermal conductivity in graphene were performed using different approaches, including the Boltzman transport equation (BTE) and molecular dynamics (MD) simulations [48-51]. The non-equilibrium phonon distribution is described by the Boltzmann transport equation [48]:

$$(\frac{\partial N_{s,\vec{q}}}{\partial t})\Big|_{drift} + (\frac{\partial N_{s,\vec{q}}}{\partial t})\Big|_{scatt} = 0, \tag{1}$$

where $N_{s,\vec{q}}$ is the number of phonons in $(s,\vec{q})$ phonon mode and $t$ is the time. The first term in Eq. (1) describes a change in $N_{s,\vec{q}}$ due to the drift motion of phonons while the second term– due to scattering. The scattering term can be rewrite within the relaxation time approximation as $(\partial N_{s,\vec{q}} / \partial t)\Big|_{Scatt} = -n_{s,\vec{q}} / \tau_{s,\vec{q}}$ , where $\tau_{s,\vec{q}}$ is the lifetime of the $(s,\vec{q})$ phonon mode, $N_0 = 1/(Exp(\hbar\omega_{(s,\vec{q})} / (k_B T)) - 1)$ is the Bose-Einstein distribution function, $T$ is the temperature and $k_B$ is the Boltzmann constant. The latter leads to the following equation:

$$n = \tau(\frac{\partial N}{\partial t})\Big|_{drift} = -\tau(\vec{v}\nabla T)\frac{\partial N_0}{\partial T}, \tag{2}$$

where $\nabla T$ is the temperature gradient and $\vec{v} = \partial \omega / \partial \vec{q}$ is the phonon group velocity. The heat flux along a graphene flake is given by the following expression [48, 52]

$$\vec{W} = \sum_{s,\vec{q}} \vec{v}_{s,\vec{q}} \hbar \omega_{s,\vec{q}} n_{s,\vec{q}}. \qquad (3)$$

Substituting Eq. (2) into Eq. (3) one has

$$\vec{W} = -\sum_{\beta=x,y,z} (\nabla T)_\beta \sum_{s,\vec{q}} \tau_{s,\vec{q}} (v_{s,\vec{q}})_\beta \frac{\partial N_0(\omega_{s,\vec{q}})}{\partial T} \vec{v}_{s,\vec{q}} \hbar \omega_{s,\vec{q}}. \qquad (4)$$

In the macroscopic approach, the thermal conductivity tensor $\kappa_{\alpha\beta}$ is defined by the equation:

$$W_\alpha = -\kappa_{\alpha\beta} (\nabla T)_\beta h L_x L_y, \qquad (5)$$

where $L_x$ and $L_y$ are the in-plane dimensions of the graphene flake and $h$ is the flake thickness. From Eqs. (4) and (5) one can obtain:

$$\kappa_{\alpha\beta} = \frac{1}{h L_x L_y} \sum_{s,\vec{q}} \tau_{s,\vec{q}} (v_{s,\vec{q}})_\alpha (v_{s,\vec{q}})_\beta \frac{\partial N_0(\omega_{s,\vec{q}})}{\partial T} \hbar \omega_{s,\vec{q}}. \qquad (6)$$

Making a transition from the summation over all phonon modes to the integration over the phonon wave vector and taking into account two-dimension phonon density of states, one can rewrite Eq. (6) for the scalar thermal conductivity $\kappa \equiv \kappa_{\alpha\alpha}$:

$$\kappa = \frac{1}{4\pi k_B T^2 h} \sum_{s=1..6} \int_0^{q_{max}} \{ [\hbar \omega_{s,\vec{q}} \frac{\partial \omega_{s,\vec{q}}}{\partial q}]^2 \tau_{s,\vec{q}} \frac{exp(\hbar \omega_{s,\vec{q}} / k_B T)}{[exp(\hbar \omega_{s,\vec{q}} / k_B T) - 1]^2} q \} dq. \qquad (7)$$

In Eq. (7), the summation $s = 1, 2, ..., 6$ is performed over six phonon branches in graphene: four in plane branches - longitudinal acoustic (LA), longitudinal optic (LO), transverse acoustic (TA), transverse optic (TO) and two out-of-plane branches – transverse acoustic (ZA) and transverse optic (ZO). The calculation of $\kappa$ requires knowledge about the mechanisms of phonon scattering in graphene. The following types of phonon scattering are usually taken into account [17, 48-51, 53]: three-phonon Umklapp and normal scatterings, edge scattering, impurity, point-defect and isotope scatterings.

Using BTE or MD approaches the RT thermal conductivity values in a range from 100 to 8000 W/mK were predicted in dependence on the edge quality, isotope and defect concentration, flake shape and size [48 - 51]. The unusual strong influence of extrinsic parameters on the thermal conductivity in graphene was explained by the relatively weak phonon – phonon scattering, resulting in a large value of the intrinsic phonon mean free path ~ 775 nm [27]. Additional scattering of phonons on the flake edges or defects decreases significantly the phonon mean free path even for small defect concentration or smooth edges. The latter opens up a promising possibility for the fine tuning the phonon heat conduction in graphene by changing structural or geometric parameters of graphene flakes. Detailed description of thermal conductivity in graphene and comparison of the results from different groups was provided in recent reviews [35, 53-54].

### III. Modeling-based design of graphene heat spreaders

Following the discovery of graphene's extremely high thermal conductivity, graphene was proposed as candidate material for heat removal applications [1, 7, 55]. Early-stage works were focused on the design of graphene heat spreaders based on modeling results [56-58]. Graphene heat spreaders were designed in different device structures and heat propagation was simulated via finite element analysis method. The comparison of temperature rise in given device structures with and without graphene heat spreaders can illustrate the efficiency of graphene heat spreaders for the improvement of heat removal capability. The feasibility study of the use of graphene as the material for lateral heat spreaders in silicon-on-insulator (SOI)-based chips was reported in Ref. [56]. Figure 3 (a) shows the design of graphene lateral heat spreaders in SOI integrated circuits. A graphene heat spreader layer was sandwiched between the oxide layer and the Si substrate and the two ends of graphene layer were connected to heat sinks. A conventional heat sink was attached to the bottom of Si substrate. The heat source was presented with several identical rectangular boxes of uniform power density, which simulated the heat generated by multiple metal-oxide-semiconductor field effect transistors (MOSFETs) in SOI circuits. The simulated temperature profiles along the top surface of a SOI based MOSFET with (black) and without (red) graphene heat spreaders are shown in Figure 3 (b). The insert shows the temperature profile for SOI wafer with seven active MOSFET devices. For a given device structure and power density, these simulations suggest that the hot-spot temperature can be reduced down to 70 K when graphene and FLG layers are embedded in the chip. The effect of graphene lateral heat spreader was more pronounced when the number of active transistors increases. It was also suggested that FLG heat spreaders may be more technologically feasible than signal layer graphene heat spreaders.

The design of graphene heat spreaders and interconnects in three-dimension (3D) integrated circuits have also been reported [55, 57]. Graphene layers incorporated into 3D chips can help in spreading heat laterally and cooling hot spots generated by Joule heating. Figure 4 (a) and (b) show the schematic of proposed 3D chip design with imbedded graphene layers as heat spreaders. The studied 3D chip contains two strata, each of which consists of a device layer and two interconnect layers. The main heat sink locates at the bottom of the substrate and additional heat sinks are connected to the ends of graphene heat spreaders. The vertical thermal via is also included in the chip design. Heat propagation equation was solved numerically using the finite element method. The simulated temperature profile of designed 3D chip are shown in Figure 4 (c) without and (d) with graphene heat spreaders. The hot spot temperature in 3D chip with imbedded graphene heat spreaders (~393 K) is substantially lower than that without graphene heat spreaders (~446 K).

### IV. Experimental demonstration of graphene heat spreaders for high-power transistors

The first experimental feasibility study of graphene lateral heat spreaders for electronic devices was demonstrated on GaN FETs [59]. The high-power GaN FETs are attractive for high-frequency high-power applications [60-62]. Commercial AlGaN/GaN heterostructure field-effect transistors (HFETs) emerged in 2005 and have developed rapidly since then. They have been used as power amplifiers or switches in wireless communications, power grids, radars, electric cars etc. AlGaN/GaN HFETs possess high electrical breakdown voltage [63], allowing high drain voltage applied. The large charge carrier concentration and saturation velocity lead to high saturation current. Therefore input power density of AlGaN/GaN HFETs could be extremely high, resulting in high output power [64-66]. Amplifiers fabricated using AlGaN/GaN HFETs have produced RF power over a wide frequency range up to several hundred watts, that is an order of magnitude larger than GaAs or InP based power devices. However, such high power density inevitably leads to huge amount of heat generation and presents extreme heat dissipation demands. Temperature rise due to self heating might lead to severe performance degradation and reliability issues [7, 67, 68]. Performance degradation of GaN transistors at elevated operation temperature includes degradation of drain current, gain and output power, as well as an increase in the gate leakage current. Moreover, the mean time to failure of GaN FETs decreases exponentially with increasing operation temperature [7]. For commercial AlGaN/GaN HFETs, the required lifetime is around $10^6$ hours and the corresponding operation temperature is below 180 $^o$C. Improvement of heat removal capability can reduce thermal resistance and increase the output power and lifetime of AlGaN/GaN HFETs. Similar considerations apply to GaN-based LEDs used in solid-state lighting applications.

It has been demonstrated [59] that the local thermal management of AlGaN/GaN HFETs can be substantially improved via introduction of the top-surface FLG heat spreaders. In the proof-of-concept experiments, FLG films have ben exfoliated from the highly-oriented pyrolytic graphite (HOPG) and transferred to the AlGaN/GaN devices on SiC substrate using polymethyl methacrylate (PMMA) membrane as supporting material. The method was analogous to the one used for transfer of the mechanically exfoliated graphene flake onto a boron nitride (BN) substrate [69]. The method was modified to allow for a fast transfer with the accuracy of spatial alignment around 1~2 μm. Figure 5 illustrates the structures of tested AlGaN/GaN HFETs and the schematics of FLG flakes transferred on top of it as top-surface heat spreaders. The tested AlGaN/GaN HFETs consisted of 30 nm AlGaN (~20% Al) barrier on 0.5 μm thick GaN channel layer deposited on insulating 4H-SiC substrate. The source and drain metal contacts were made of Ti/Al/Ti/Au, while the gate electrode was made of Ni/Au. The gate length and widths of the devices were 3.5 and 90 μm, respectively. The large source drain separation of 12 μm facilitated the heat spreader fabrication. The mobility values for representative devices were around 1150 cm$^2$/Vs.

Figure 6 provides microscopy images of graphene heat spreaders transferred on top of AlGaN/GaN HFETs. Since graphene film is electrically conductive, in order to avoid short circuiting the tested device, the graphene heat spreaders extend from the drain contact directly to the heat sinks on the side of the device. Microscopy images also show the flexibility of graphene heat spreaders and close contact between graphene and the sample surface. The performance of graphene heat spreaders was demonstrated by comparison of temperature rise in operating AlGaN/GaN HFETs at same dissipation power with and without heat spreaders. The temperature rise in the device channel was in-situ monitored via Raman spectroscopy [70-72]. Raman spectrum of the FLG on AlGaN/GaN/SiC layered structure shows characteristic Raman peaks of FLG, GaN and SiC [59]. The narrow Ramam peak at 567 cm$^{-1}$ is $E_2$ mode of GaN. That peak position is sensitive to temperature and the temperature dependence has been well established [73,74], thus GaN $E_2$ peak can be utilized for temperature measurement.

AlGaN/GaN HFET with graphene heat spreaders and the reference HFET without the heat spreaders were wire-bonded and placed under the Raman microscope (Renishaw inVia Raman system). Direct current (DC) bias was applied to the tested devices and temperature rise, $\Delta T$, due to self-heating in the device channel was monitored by the Raman peak positions. Figure 7 shows GaN $E_2$ peak in the Raman spectra of two identical AlGaN/GaN devices with and without graphene heat spreaders. The laser spot was focused at the channel region between the gate and the drain, closer to the gate, where $\Delta T$ is expected to be the highest. At a power density of 12.8 W/mm, the temperature rise, $\Delta T$, for the AlGaN/GaN HFET with and without graphene heat spreaders was 92 °C and 118 °C, respectively. In that measurement, same dissipation power was achieved in the latter device at 22 V source-drain bias due to small variations in the current–voltage characteristics (I-V). The results proved that graphene heat spreaders reduced the hotspot temperature around 20 °C of the tested device at given power density.

Figure 8 provides direct comparison of I-V characteristics of the HFETs with (solid lines) and without (dashed lines) graphene heat spreaders. At $V_G = 2$ V, $I_{SD}$ increases from ~0.75 A/mm to ~0.84 A/mm – 12 % improvement – as a result of better heat removal with the top lateral heat spreaders. At $V_G = 0$ V, $I_{SD}$ increased from 0.47 A/mm to 0.51 A/mm, which is 8 % improvement. At $V_G = -2$V, the current density remains almost the same due to the low dissipation power density at this negative gate bias. Those experiments presented a direct evidence of the improvement in the AlGaN/GaN HFET performance with the top-surface few-layer graphene heat spreaders.

To emphasize the technological importance of findings of Ref. [59] it is illustrating to compare the thermal properties of FLG with those of metals, which can also be used as heat spreaders. It is well known that the thermal conductivity of metal films rapidly decreases with the film thickness [75-77]. For many technologically important metals, e.g. aluminum, copper or gold, the thermal conductivity of the metal film, $\kappa_F$, constitutes only ~20% of the thermal conductivity of bulk metal, $\kappa_M$, at the film thickness $H \approx 100$ nm. For example, the thermal conductivity of the gold film on etched Si for $H$ approaching the electron mean free path $\lambda \approx 41$ nm is $\kappa_F \approx 0.2 \times \kappa_M$ [77]. The expected down-scaling for aluminum films would give $\kappa_F \approx 26 - 48$ W/mK considering that the bulk RT $\kappa_M$ value for aluminum ranges from ~130 W/mK to 240 W/mK, depending on its purity and quality. The drastic degradation of the heat conduction properties of metal films is due to the increased electron scattering from the rough surfaces of the films and the polycrystalline grain boundaries. The surface roughness of thin metal films is usually rather high leading to stronger diffusive phonon scattering from interfaces. From the other side thermal conductivity of FLG is close to the bulk graphite limit of ~2000 W/mK and can be even larger up to 4000 W/mK for n<4. For this reason, the thermal conductivity of FLG is larger than that of thin metal films almost *by* *two orders of magnitude* leading to substantial differences in the heat fluxes when these materials are used as heat spreaders.

## V. CVD grown graphene heat spreaders

The first experimental demonstration of graphene heat spreaders on high-power electronic device was achieved by transferring mechanical exfoliated FLG on GaN transistors. However, this method cannot be applied in semiconductor industry due to the limited graphene flake size, randomness of flake shape and thickness, as well as the low throughput. The practical applications of graphene heat spreaders will rely on a method that could produce large size graphene flakes of high quality at low price. Fast progress of graphene growth by CVD method [78-80] and other techniques can make this feasible in the near future.

The reported thermal conductivity of CVD grown graphene is lower than that of exfoliated graphene flakes [36, 81], but still larger than conventional semiconductors or metals used in electronic devices. In a recent work, CVD grown graphene of different number of layers was fabricated and it was demonstrated as heat spreader in thermal packaging [82]. A platinum microheater embedded chips were used to evaluate the performance of the graphene heat spreaders. The Pt microheater made of titanium/platinum/gold (Ti/Pt/Au) provided heating source and temperature sensor. CVD grown single-layer and multilayer graphene were synthesized on copper substrate and then transferred onto the thermal evaluation chips as heat spreaders. Pt microheater was driven by electric current as hot spot, in which the temperature rise can be calculated by measuring the electric resistance. Thermal performance of the graphene heat spreaders was evaluated by the temperature drop after graphene transfer. It was found that the temperature of hot spot driven at a heat flux up to 430 Wcm$^{-2}$ was decreased from 121 $^o$C to 108 $^o$C ($\varDelta T = 13$ $^o$C) after introduction of SLG heat spreader. These results prove the potential of CVD grown graphene as a promising heat spreader material for hot spot cooling in electronic devices.

## VI. Conclusions

We reviewed thermal properties of graphene and multilayer graphene, and discussed possible graphene applications in heat spreaders for thermal management of high-power electronic and optoelectronic devices. Graphene heat spreaders can efficiently improve heat removal owing to the high thermal conductivity of graphene and its compatibility with various substrate materials. In addition, compared with conventional nanometer-thick thin films or nanowires, graphene thin films or graphene nanoribbons can maintain the high thermal conductivity at nanometer scale. The latter is important for the device-level targeted cooling of micrometer or nanometer scale hotspots. The proposed local heat spreading with materials that preserve good thermal properties at nanometre scale represents a transformative change in thermal management.

## Acknowledgments

AAB acknowledges funding from the Semiconductor Research Corporation (SRC) and Defense Advanced Research Project Agency (DARPA). NDL acknowledges the financial support from the Republic of Moldova through the projects 11.817.05.10F and 14.819.16.02F and from the Science and Technology Center in Ukraine (STCU, project #5937).

**References**

[1]  A. A. Balandin: 'Chill Out', *IEEE Spectrum*, 2009, vol. 46, pp. 34-39

[2]  S. V. Garimella, A. S. Fleischer, J. Y. Murthy*, et al.*: 'Thermal Challenges in Next-Generation Electronic Systems', *IEEE Transactions on Components and Packaging Technologies,* , 2008, vol. 31, pp. 801-815

[3]  R. Prasher: 'Thermal Interface Materials: Historical Perspective, Status, and Future Directions', *Proceedings of the IEEE*, 2006, vol. 94, pp. 1571-1586

[4]  S. Farhad, D. C. Whalley, and P. P. Conway, "Thermal Interface Materials - A Review of the State of the Art," in *Electronics Systemintegration Technology Conference, 2006. 1st*, 2006, pp. 1292-1302.

[5]  K. Puttaswamy and G. H. Loh, "Thermal analysis of a 3D die-stacked high-performance microprocessor," presented at the Proceedings of the 16th ACM Great Lakes symposium on VLSI, Philadelphia, PA, USA, 2006.

[6]  A. Christensen and S. Graham: 'Thermal effects in packaging high power light emitting diode arrays', *Applied Thermal Engineering*, 2009, vol. 29, pp. 364-371

[7]  G. Meneghesso, G. Verzellesi, F. Danesin*, et al.*: 'Reliability of GaN High-Electron-Mobility Transistors: State of the Art and Perspectives', *Device and Materials Reliability, IEEE Transactions on*, 2008, vol. 8, pp. 332-343

[8]  M. Meneghini, L. R. Trevisanello, G. Meneghesso, and E. Zanoni: 'A Review on the Reliability of GaN-Based LEDs', *Device and Materials Reliability, IEEE Transactions on*, 2008, vol. 8, pp. 323-331

[9]  V. O. Turin and A. A. Balandin: 'Electrothermal simulation of the self-heating effects in GaN-based field-effect transistors', *Journal of Applied Physics*, 2006, vol. 100, pp. 054501-054501-8

[10]  A. Sarua, H. Ji, M. Kuball*, et al.*: 'Integrated micro-Raman/infrared thermography probe for monitoring of self-heating in AlGaN/GaN transistor structures', *Electron Devices, IEEE Transactions on*, 2006, vol. 53, pp. 2438-2447

[11]  M. A. Stroscio and M. Dutta, *Phonons in Nanostructures*. Cambridge, United Kingdom: Cambridge University Press, 2001.

[12]  A. Balandin and K. L. Wang: 'Effect of phonon confinement on the thermoelectric figure of merit of quantum wells', *Journal of Applied Physics*, 1998, vol. 84, pp. 6149-6153

[13]   A. Balandin and K. L. Wang: 'Significant decrease of the lattice thermal conductivity due to phonon confinement in a free-standing semiconductor quantum well', *Physical Review B*, 1998, vol. 58, pp. 1544-1549

[14]   D. V. Crismari and D. L. Nika: 'Thermal Conductivity Reduction in Si/Ge Core/Shell Nanowires', *Journal of Nanoelectronics and Optoelectronics*, 2012, vol. 7, pp. 701-705

[15]   E. P. Pokatilov, D. L. Nika, and A. A. Balandin: 'A phonon depletion effect in ultrathin heterostructures with acoustically mismatched layers', *Applied Physics Letters*, 2004, vol. 85, pp. 825-827

[16]   D.L. Nika, E.P. Pokatilov and A.A. Balandin: 'Phonon-engineered mobility enhancement in the acoustically mismatched silicon/diamond transistor channels', *Applied Physics Letters*, 2008, vol. 93, p. 173111

[17]   A.A. Balandin and D.L. Nika: 'Phononics in low-dimensional materials', *Materials Today*, 2012, vol. 15, pp. 266-275

[18]   N.D. Zincenco, D.L. Nika, E.P. Pokatilov, and A.A. Balandin: 'Acoustic phonon engineering of thermal properties of silicon-based nanostructures', Journal of Physics: Conference Series, 2007, vol. 92, p. 012022

[19]   W. Liu and M. Asheghi: 'Thermal Conductivity Measurements of Ultra-Thin Single Crystal Silicon Layers', *Journal of Heat Transfer*, 2005, vol. 128, pp. 75-83

[20]   D. Li, Y. Wu, P. Kim, L. Shi, P. Yang, and A. Majumdar: 'Thermal conductivity of individual silicon nanowires', *Applied Physics Letters*, 2003, vol. 83, pp. 2934-2936

[21]   A. I. Hochbaum, R. Chen, R. D. Delgado, *et al.*: 'Enhanced thermoelectric performance of rough silicon nanowires', *Nature*, 2008, vol. 451, pp. 163-167

[22]   P. Martin, Z. Aksamija, E. Pop, and U. Ravaioli: 'Impact of Phonon-Surface Roughness Scattering on Thermal Conductivity of Thin Si Nanowires', *Physical Review Letters*, 2009, vol. 102, p. 125503

[23]   D.L. Nika, E.P. Pokatilov, A. A. Balandin, *et al.*: 'Reduction of lattice thermal conductivity in one-dimensional quantum-dot superlattice due to phonon filtering', *Physical Review B*, 2011, vol. 84, p. 165415

[24]   D.L. Nika, A.I. Cocemasov, C.I. Isacova, *et al.*: 'Suppression of phonon heat conduction in cross-section-modulated nanowires', *Physical Review B,* 2012, vol. 85, p. 205439

[25]   D.L. Nika, A.I. Cocemasov, D.V. Crismari, *et al.*: 'Thermal conductivity inhibition in phonon engineered core-shell cross-section-modulated Si/Ge nanowires', *Applied Physics Letters*, 2013, vol. 102, p. 213109

[26]   A. A. Balandin, S. Ghosh, W. Bao, *et al.*: 'Superior Thermal Conductivity of Single-Layer Graphene', *Nano Letters*, 2008, vol. 8, pp. 902-907

[27]   S. Ghosh, I. Calizo, D. Teweldebrhan, *et al.*: 'Extremely high thermal conductivity of graphene: Prospects for thermal management applications in nanoelectronic circuits', *Applied Physics Letters*, 2008, vol. 92, p. 151911

[28]   R. R. Nair, P. Blake, A. N. Grigorenko, *et al.*: 'Fine Structure Constant Defines Visual Transparency of Graphene', *Science*, 2008, vol. 320, p. 1308

[29]   K. S. Novoselov, A. K. Geim, S. V. Morozov, *et al.*: 'Two-dimensional gas of massless Dirac fermions in graphene', *Nature*, 2005, vol. 438, pp. 197-200

[30]   Y. Zhang, Y.-W. Tan, H. L. Stormer, and P. Kim: 'Experimental observation of the quantum Hall effect and Berry's phase in graphene', *Nature*, 2005, vol. 438, pp. 201-204

[31]   A. C. Ferrari, J. C. Meyer, V. Scardaci, *et al.*: 'Raman Spectrum of Graphene and Graphene Layers', *Physical Review Letters*, 2006, vol. 97, p. 187401

[32]   I. Calizo, W. Bao, F. Miao, C. N. Lau, and A. A. Balandin: 'The effect of substrates on the Raman spectrum of graphene: Graphene- on-sapphire and graphene-on-glass', *Applied Physics Letters*, 2007, vol. 91, p. 201904

[33]   I. Calizo, F. Miao, W. Bao, C. N. Lau, and A. A. Balandin: 'Variable temperature Raman microscopy as a nanometrology tool for graphene layers and graphene-based devices', *Applied Physics Letters*, 2007, vol. 91, p. 071913

[34]   I. Calizo, A. A. Balandin, W. Bao, F. Miao, and C. N. Lau: 'Temperature Dependence of the Raman Spectra of Graphene and Graphene Multilayers', *Nano Letters*, 2007, vol. 7, pp. 2645-2649

[35]  A. A. Balandin: 'Thermal properties of graphene and nanostructured carbon materials', *Nature Materials*, 2011, vol. 10, pp. 569-581

[36]   W. Cai, A. L. Moore, Y. Zhu, *et al.*: 'Thermal Transport in Suspended and Supported Monolayer Graphene Grown by Chemical Vapor Deposition', *Nano Letters*, 2010, vol. 10, pp. 1645-1651

[37] M.E. Pumarol , M.C. Rosamond, P. Tovee, M.C. Petty, D.A. Zeze, V. Falko, and O.V. Kolosov, "Direct nanoscale imaging of ballistic and diffusive thermal transport in graphene nanostructures, *Nano Letters*, 2012, vol. 12, pp 2906–2911

[38] J.-U. Lee, D. Yoon, H. Kim, S. W. Lee, and H. Cheong: 'Thermal conductivity of suspended pristine graphene measured by Raman spectroscopy', *Physical Review B*, 2011, vol. 83, p. 081419

[39] A. B. Kuzmenko, E. van Heumen, F. Carbone, and D. van der Marel: 'Universal Optical Conductance of Graphite', *Physical Review Letters*, 2008, vol. 100, p. 117401

[40] K. F. Mak, J. Shan, and T. F. Heinz: 'Seeing Many-Body Effects in Single- and Few-Layer Graphene: Observation of Two-Dimensional Saddle-Point Excitons', *Physical Review Letters*, 2011, vol. 106, p. 046401

[41] K. S. Kim, Y. Zhao, H. Jang*, et al.*: 'Large-scale pattern growth of graphene films for stretchable transparent electrodes', *Nature*, 2009, vol. 457, pp. 706-710

[42] V. G. Kravets, A. N. Grigorenko, R. R. Nair*, et al.*: 'Spectroscopic ellipsometry of graphene and an exciton-shifted van Hove peak in absorption', *Physical Review B*, 2010, vol. 81, p. 155413

[43] S. Ghosh, W. Bao, D. L. Nika*, et al.*: 'Dimensional crossover of thermal transport in few-layer graphene', *Nat Mater*, 2010, vol. 9, pp. 555-558

[44] J. H. Seol, I. Jo, A. L. Moore*, et al.*: 'Two-Dimensional Phonon Transport in Supported Graphene', *Science*, 2010, vol. 328, pp. 213-216

[45] W. Jang, Z. Chen, W. Bao, C. N. Lau, and C. Dames: 'Thickness-Dependent Thermal Conductivity of Encased Graphene and Ultrathin Graphite', *Nano Letters*, 2010, vol. 10, pp. 3909-3913

[46] R. Murali, Y. Yang, K. Brenner, T. Beck, and J. D. Meindl: 'Breakdown current density of graphene nanoribbons', *Applied Physics Letters*, 2009, vol. 94, p. 243114.

[47] A. D. Liao, J. Z. Wu, X. Wang*, et al.*: 'Thermally Limited Current Carrying Ability of Graphene Nanoribbons', *Physical Review Letters*, 2011, vol. 106, p. 256801

[48] D.L. Nika, E.P. Pokatilov, A.S. Askerov, and A.A. Balandin: 'Phonon thermal conduction in graphene: role of Umklapp and edge roughness scattering', *Physical Review B*, 2009, vol 79, p. 155413.

[49] L. Lindsay, D.A. Broido, and N. Mingo: 'Flexural phonons and thermal transport in multilayer graphene and graphite', *Physical Review B,* 2011, vol. 83, p. 235428

[50] W.J. Evans, L. Hu, and P. Keblinsky: 'Thermal conductivity of graphene ribbons from equilibrium molecular dynamics: effect of ribbon width, edge roughness, and hydrogen termination', *Applied Physics Lett*ers, 2010, vol. 96, p. 203112

[51] D.L. Nika, A.S. Askerov, and A.A Balandin: 'Anomalous size dependence of the thermal conductivity of graphene ribbons', *Nano Letters,* 2012, vol. 12, pp. 3238 - 3244

[52] M. Ziman, *Electrons and Phonons* (Clarendon Press, Oxford, 2001).

[53] D.L. Nika and A.A. Balandin: Two-dimensional phonon transport in graphene', *Journal of Physics: Condensed Matter,* 2012, vol. 24, p. 233203

[54] P. Goli, D.L. Nika and A.A. Balandin: ' Graphene Thermal Properties and Applications in Thermal Phase Change Materials', Applied Sciences (submitted, 2014)

[55] A. A. Balandin: 'The Heat Is On: Graphene Applications', *Nanotechnology Magazine, IEEE*, 2011, vol. 5, pp. 15-19

[56] S. Subrina, D. Kotchetkov, and A. A. Balandin: 'Heat Removal in Silicon-on-Insulator Integrated Circuits With Graphene Lateral Heat Spreaders', *Electron Device Letters, IEEE*, 2009, vol. 30, pp. 1281-1283

[57] S. Subrina: 'Modeling Based Design of Graphene Heat Spreaders and Interconnects in 3-D Integrated Circuits', *Journal of Nanoelectronics and Optoelectronics*, 2010, vol. 5, pp. 281-286

[58] S. Subrina, D. Kotchetkov, and A. A. Balandin, "Thermal management with graphene lateral heat spreaders: A feasibility study," in *Thermal and Thermomechanical Phenomena in Electronic Systems (ITherm), 2010 12th IEEE Intersociety Conference on*, 2010, pp. 1-5.

[59] Z. Yan, G. Liu, J. M. Khan, and A. A. Balandin: 'Graphene quilts for thermal management of high-power GaN transistors', *Nat Commun*, 2012, vol. 3, p. 827

[60] M. Asif Khan, A. Bhattarai, J. N. Kuznia, and D. T. Olson: 'High electron mobility transistor based on a GaN‐$Al_xGa_{1-x}N$ heterojunction', *Applied Physics Letters*, 1993, vol. 63, pp. 1214-1215

[61] U. K. Mishra, P. Parikh, and W. Yi-Feng: 'AlGaN/GaN HEMTs-an overview of device operation and applications', *Proceedings of the IEEE*, 2002, vol. 90, pp. 1022-1031

[62]    M. S. Shur: 'GaN based transistors for high power applications', *Solid-State Electronics*, 1998, vol. 42, pp. 2131-2138

[63]    J. Li, S. J. Cai, G. Z. Pan, Y. L. Chen, C. P. Wen, and K. L. Wang: 'High breakdown voltage GaN HFET with field plate', *Electronics Letters*, 2001, vol. 37, pp. 196-197

[64]    S. T. Sheppard, K. Doverspike, W. L. Pribble*, et al.*: 'High-power microwave GaN/AlGaN HEMTs on semi-insulating silicon carbide substrates', *Electron Device Letters, IEEE*, 1999, vol. 20, pp. 161-163

[65]    L. Shen, S. Heikman, B. Moran*, et al.*: 'AlGaN/AlN/GaN high-power microwave HEMT', *Electron Device Letters, IEEE*, 2001, vol. 22, pp. 457-459

[66]    W. Yi-Feng, D. Kapolnek, J. P. Ibbetson, P. Parikh, B. P. Keller, and U. K. Mishra: 'Very-high power density AlGaN/GaN HEMTs', *Electron Devices, IEEE Transactions on*, 2001, vol. 48, pp. 586-590

[67]    J. A. del Alamo and J. Joh: 'GaN HEMT reliability', *Microelectronics Reliability*, 2009, vol. 49, pp. 1200-1206

[68]    V. O. Turin and A. A. Balandin: 'Performance degradation of GaN field-effect transistors due to thermal boundary resistance at GaN/substrate interface', *Electronics Letters*, 2004, vol. 40, pp. 81-83

[69]    C. R. Dean, A. F. Young, I. Meric *et al.*: 'Boron nitride substrates for high-quality graphene electronics', *Nat Nano*, 2010, vol. 5, pp. 722-726

[70]    M. Kuball, J. M. Hayes, M. J. Uren*, et al.*: 'Measurement of temperature in active high-power AlGaN/GaN HFETs using Raman spectroscopy', *Electron Device Letters, IEEE*, 2002, vol. 23, pp. 7-9

[71]    M. Kuball, S. Rajasingam, A. Sarua*, et al.*: 'Measurement of temperature distribution in multifinger AlGaN/GaN heterostructure field-effect transistors using micro-Raman spectroscopy', *Applied Physics Letters*, 2003, vol. 82, pp. 124-126

[72]    R. J. T. Simms, J. W. Pomeroy, M. J. Uren, T. Martin, and M. Kuball: 'Channel Temperature Determination in High-Power AlGaN/GaN HFETs Using Electrical Methods and Raman Spectroscopy', *Electron Devices, IEEE Transactions on*, 2008, vol. 55, pp. 478-482

[73]     M. S. Liu, L. A. Bursill, S. Prawer, K. W. Nugent, Y. Z. Tong, and G. Y. Zhang: 'Temperature dependence of Raman scattering in single crystal GaN films', *Applied Physics Letters*, 1999, vol. 74, pp. 3125-3127

[74]     W. S. Li, Z. X. Shen, Z. C. Feng, and S. J. Chua: 'Temperature dependence of Raman scattering in hexagonal gallium nitride films', *Journal of Applied Physics*, 2000, vol. 87, pp. 3332-3337

[75]     P. Nath and K. L. Chopra: 'Thermal conductivity of copper films', *Thin Solid Films*, 1974, vol. 20, pp. 53-62

[76]     G. Chen and P. Hui: 'Thermal conductivities of evaporated gold films on silicon and glass', *Applied Physics Letters*, 1999, vol. 74, pp. 2942-2944

[77]     G. Langer, J. Hartmann, and M. Reichling: 'Thermal conductivity of thin metallic films measured by photothermal profile analysis', *Review of Scientific Instruments*, 1997, vol. 68, pp. 1510-1513

[78]     A. N. Obraztsov: 'Chemical vapour deposition: Making graphene on a large scale', *Nat Nano*, 2009, vol. 4, pp. 212-213

[79]     A. Reina, X. Jia, J. Ho*, et al.*: 'Large Area, Few-Layer Graphene Films on Arbitrary Substrates by Chemical Vapor Deposition', *Nano Letters*, 2008, vol. 9, pp. 30-35

[80]     S. Bae, H. Kim, Y. Lee*, et al.*: 'Roll-to-roll production of 30-inch graphene films for transparent electrodes', *Nat Nano*, 2010, vol. 5, pp. 574-578

[81]     Y.-H. Lee and J.-H. Lee: 'Scalable growth of free-standing graphene wafers with copper(Cu) catalyst on SiO2/Si substrate: Thermal conductivity of the wafers', *Applied Physics Letters*, 2010, vol. 96, pp. -

[82]     Z. Gao, Y. Zhang, Y. Fu, M. M. F. Yuen, and J. Liu: 'Thermal chemical vapor deposition grown graphene heat spreader for thermal management of hot spots', *Carbon*, 2013, vol. 61, pp. 342-348

**Figure Captions**

**Figure 1:** (a) Schematic of the thermal conductivity measurement showing suspended FLG flakes and excitation laser light. (b) Scanning electron microscopy image of the suspended graphene flake. Figure adopted from Ref. [35] with permission from Nature Publishing Group.

**Figure 2:** Measured thermal conductivity of suspended FLG as a function of the number of graphene layers, $n$. When $n > 4$, thermal conductivity drops below the bulk graphite limit due to the onset of phonon boundary scattering. It approaches bulk graphite limit when $n$ keeps increasing. Figure adopted from Ref. [43] with permission from Nature Publishing Group.

**Figure 3:** Graphene heat spreaders designed for SOI integrated circuits. (a) Schematic of the circuit on SOI substrate with graphene lateral heat spreaders attached to the side heat sinks and a main heat sink at the bottom of the substrate. (b) Temperature profile on the top surface of a SOI-based MOSFET with (black) and without (red) graphene heat spreaders. The insert shows the temperature profile for a SOI chip with seven active devices. Figure adopted from Ref. [55] with permission from IEEE.

**Figure 4:** Graphene heat spreaders designed for 3D integrated circuits. (a) A 3D view and (b) cross section of the schematic of designed 3D chip showing graphene heat spreaders for heat removal from the localized hot spots and interconnects. Simulated temperature profile across the 3D chip are shown for the designs (c) without and (d) with graphene heat spreaders. The units of temperature are in degree Kelvin. Figure adopted from Ref. [54] with permission from IEEE.

**Figure 5:** (a) Schematic of the FLG-graphite heat spreader attached to the AlGaN/GaN HFET (b) Schematic of the structure of tested device and graphene heat spreader. Figure adopted from Ref. [58] with permission from Nature Publishing Group.

**Figure 6:** Microscopy images of graphene heat spreaders transferred on AlGaN/GaN HFET. (a) Optical microscopy image of FLG overlapping the drain contact and the top surface of AlGaN/GaN HFET. Scale bar is 100μm. (b) Scanning electron microscopy (SEM) image of the heat spreader transferred on the top of drain contact. Graphene is indicated with green color, while metal contacts are with yellow color. Scale bar is 10μm. (c) A typical SEM image of FLG overlapping the boundary between metal contact and GaN surface. Scale bar is 1μm. Figure adopted from Ref. [58]with permission from Nature Publishing Group.

**Figure 7:** Comparison of temperature rise in operating AlGaN/GaN HFETs measured by Raman spectroscopy. (a) GaN $E_2$ peak shift in AlGaN/GaN HFET without graphene heat spreader. (b) GaN $E_2$ peak shift in AlGaN/GaN HFET with graphene heat spreader at same dissipation power. Smaller Raman peak shift indicating a temperature reduction in operating device. Figure adopted from Ref. [58] with permission from Nature Publishing Group.

**Figure 8:** Comparison of I-Vs of AlGaN/GaN HFETs with (dashed lines) and without (solid lines) graphene heat spreaders indicating improvement in I-Vs in HFETs after adding the lateral heat spreaders. Figure adopted from Ref. [58] with permission from Nature Publishing Group.

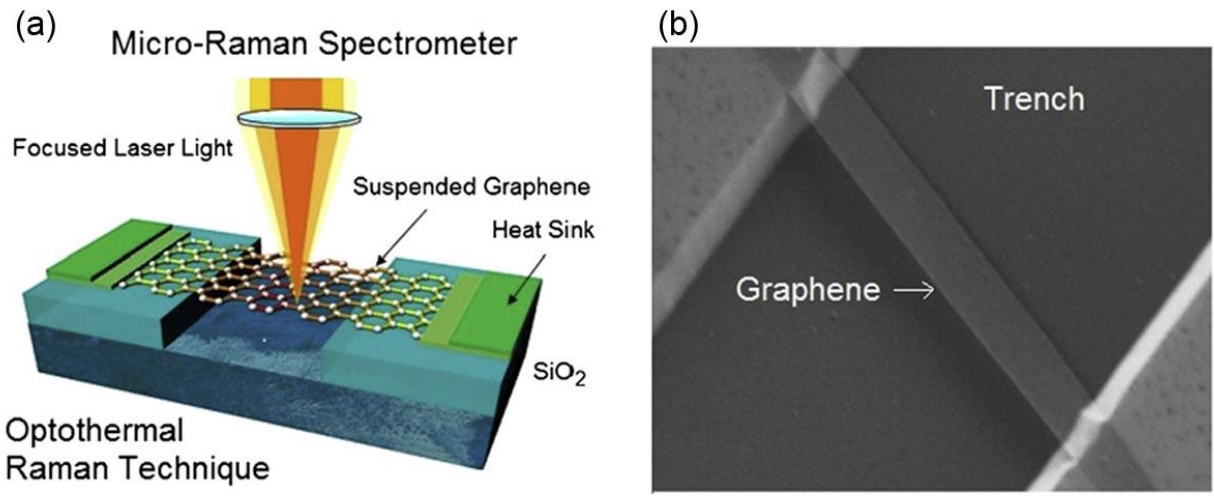

Figure 1

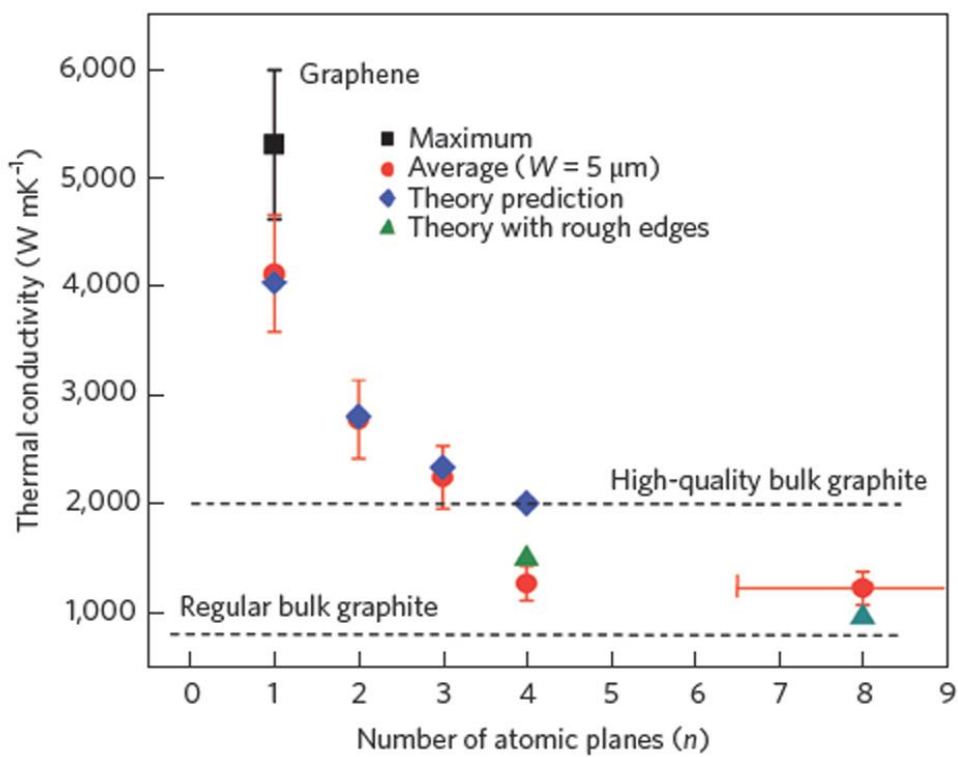

Figure 2

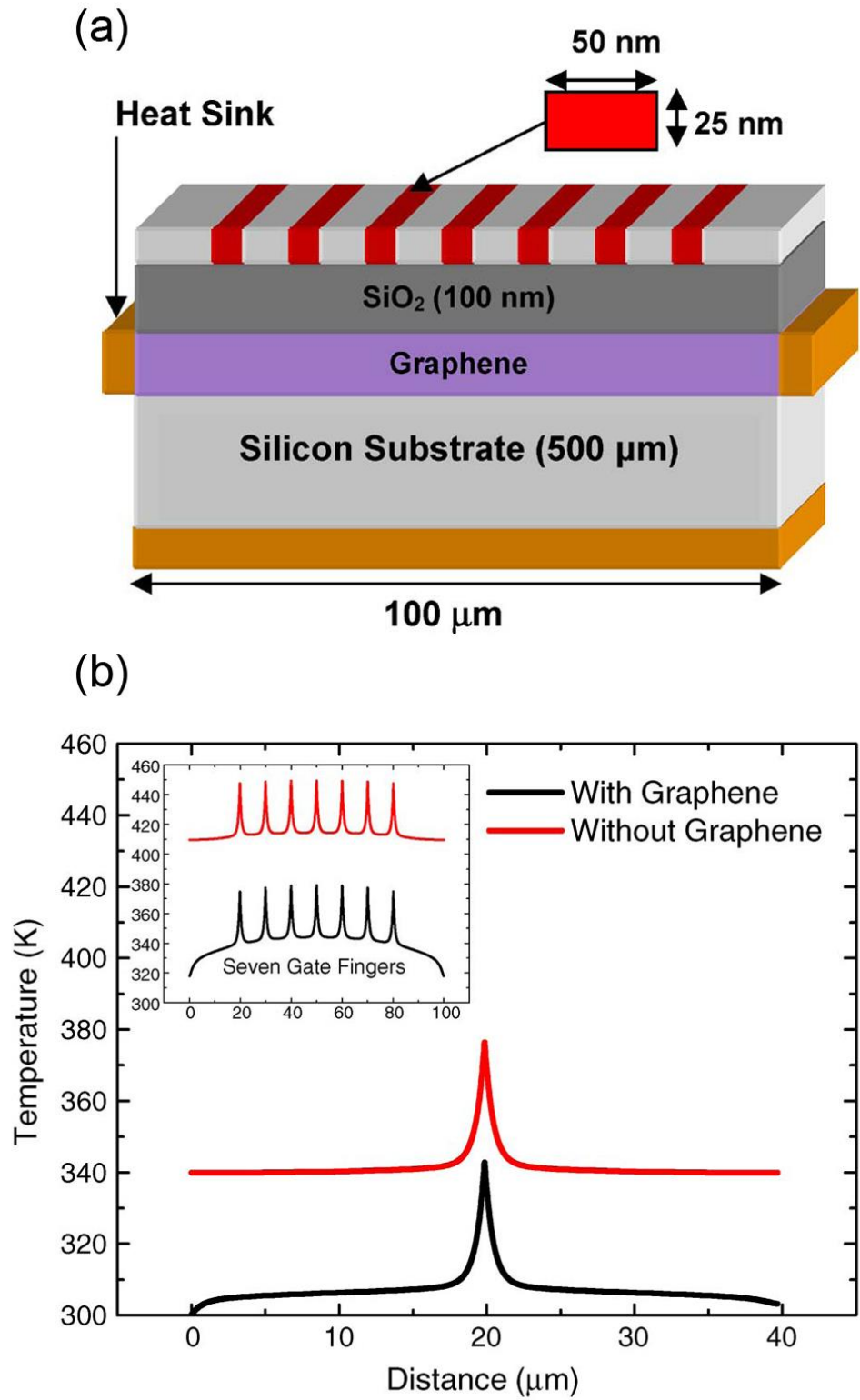

Figure 3

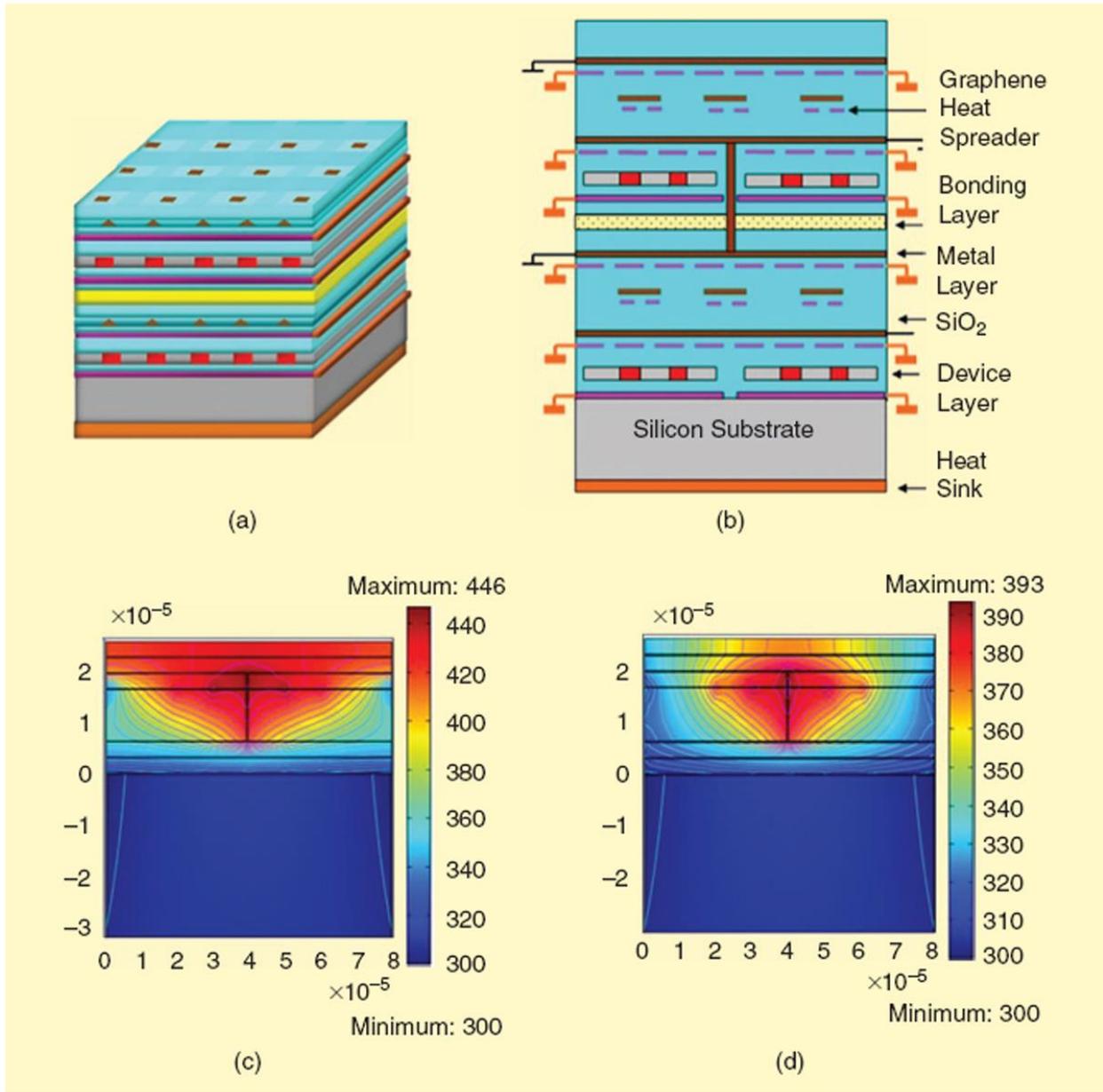

Figure 4

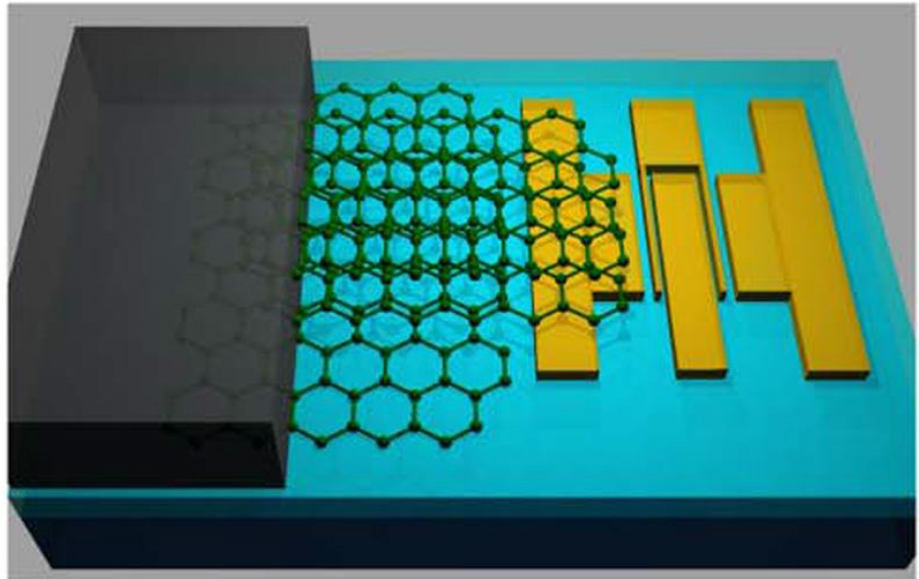

(a)

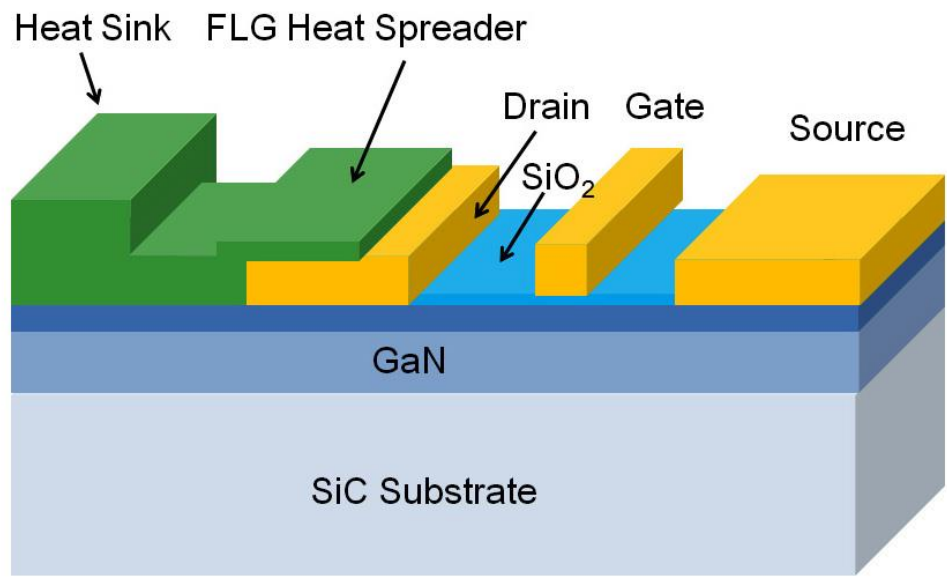

(b)

Figure 5

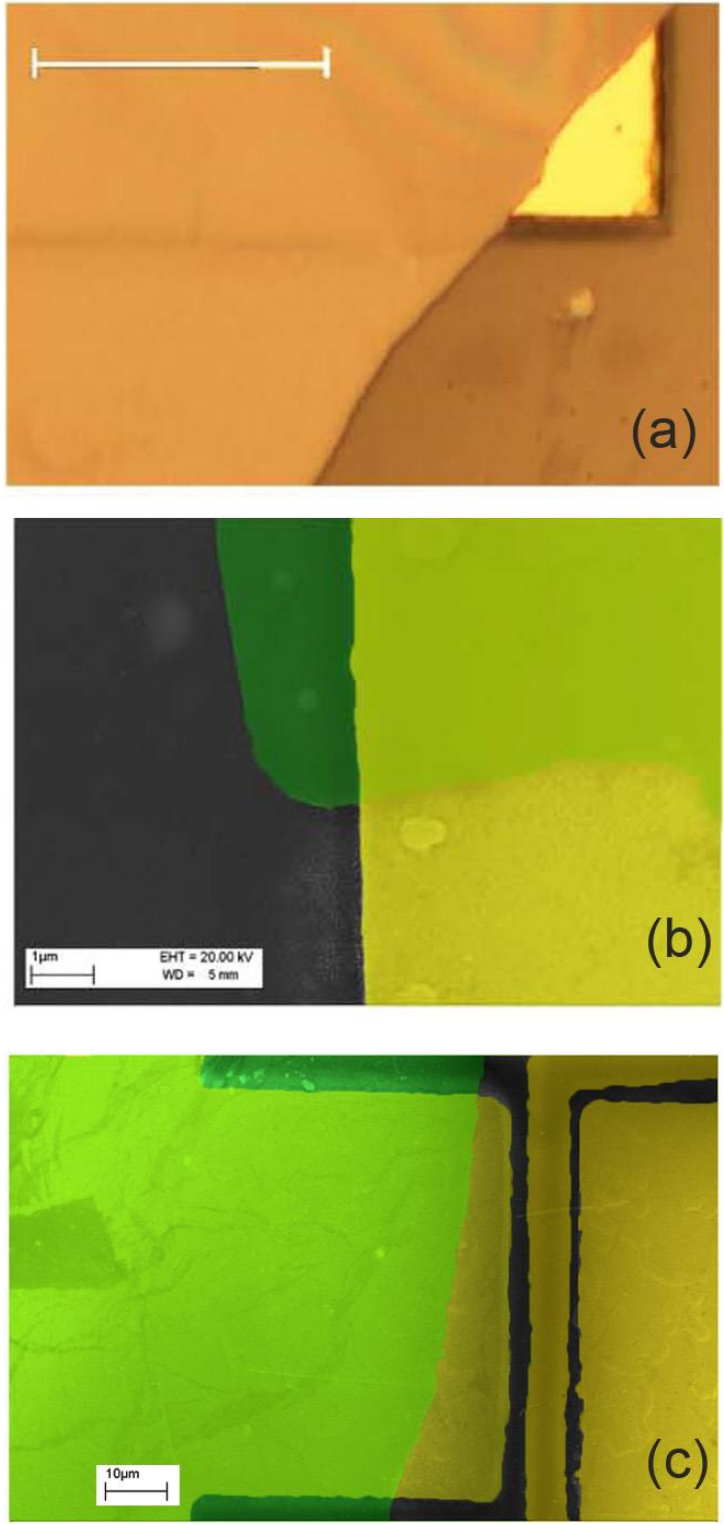

Figure 6

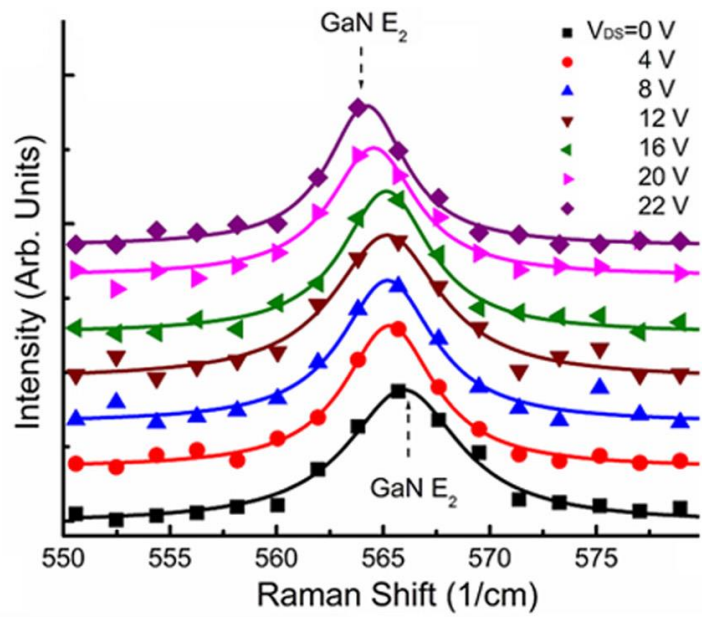

(a)

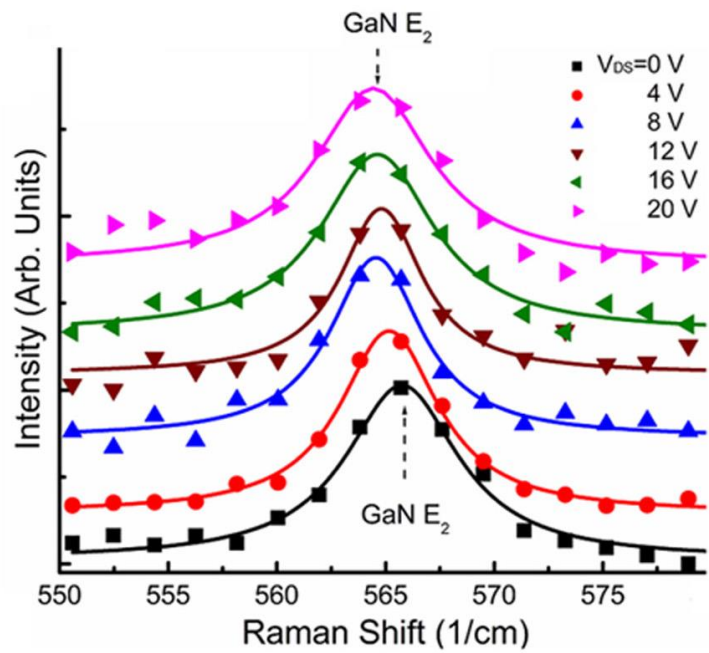

(b)

Figure 7

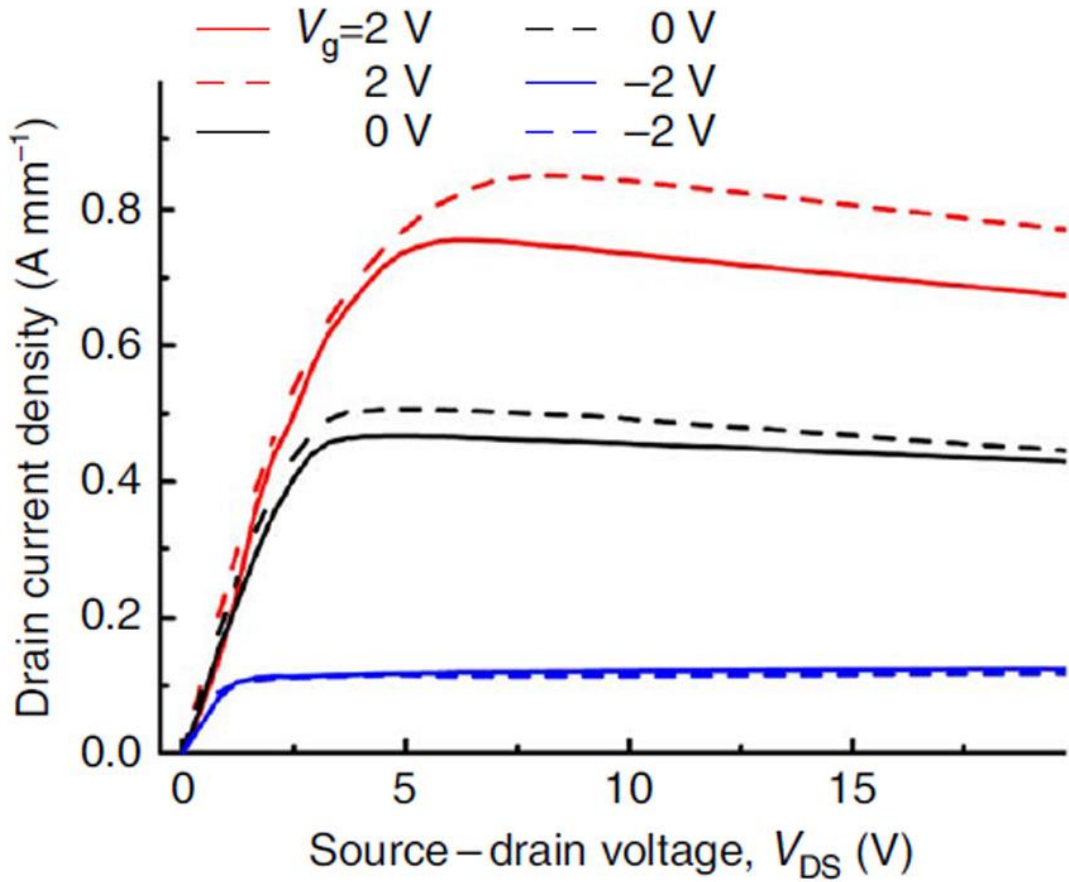

Figure 8